\documentclass[apj]{emulateapj}
\usepackage{apjfonts}
\usepackage{graphicx}

\shorttitle{Inner/Outer Halo Fractions in the Local Volume}
\shortauthors{An et~al.}

\begin{document}

\title{The Fractions of Inner- and Outer-Halo Stars in the Local Volume}

\author{Deokkeun An\altaffilmark{1},
Timothy C.\ Beers\altaffilmark{2},
Rafael M. Santucci\altaffilmark{3},
Daniela Carollo\altaffilmark{2},\\
Vinicius M.\ Placco\altaffilmark{2},
Young Sun Lee\altaffilmark{4},
Silvia Rossi\altaffilmark{3}
}

\altaffiltext{1}{Department of Science Education, Ewha Womans University, 52 Ewhayeodae-gil, Seodaemun-gu, Seoul 03760, Korea; deokkeun@ewha.ac.kr}
\altaffiltext{2}{Department of Physics and JINA Center for the Evolution of the Elements, University of Notre Dame, Notre Dame, IN 46556, USA}
\altaffiltext{3}{Departamento de Astronomia, Instituto de Astronomia, Geof\'{i}sica e Ci\^{e}ncias Atmosf\'{e}ricas, Universidade de S\~{a}o Paulo, S\~{a}o Paulo, SP 05508-900, Brazil}
\altaffiltext{4}{Department of Astronomy \& Space Science, Chungnam National University, Daejeon 34134, Republic of Korea}

\begin{abstract}

We obtain a new determination of the metallicity distribution function (MDF) of stars within $\sim5$--$10$ kpc of the Sun, based on recently improved co-adds of $ugriz$ photometry for Stripe~82 from the Sloan Digital Sky Survey. Our new estimate uses the methodology developed previously by \citeauthor{an:13} to study in situ halo stars, but is based on a factor of two larger sample than available before, with much-improved photometric errors and zero-points. The newly obtained MDF can be divided into multiple populations of halo stars, with peak metallicities at [Fe/H] $\approx -1.4$ and $-1.9$, which we associate with the inner-halo and outer-halo populations of the Milky Way, respectively. We find that the kinematics of these stars (based on proper-motion measurements at high Galactic latitude) supports the proposed dichotomy of the halo, as stars with retrograde motions in the rest frame of the Galaxy are generally more metal-poor than stars with prograde motions, consistent with previous claims. In addition, we generate mock catalogs of stars from a simulated Milk Way halo system, and demonstrate for the first time that the chemically- and kinematically-distinct properties of the inner- and outer-halo populations are qualitatively in agreement with our observations. The decomposition of the observed MDF and our comparison with the mock catalog results suggest that the outer-halo population contributes on the order of $\sim35\%$--$55\%$ of halo stars in the local volume.
    
\end{abstract}


\section{Introduction}

While many details of the galaxy-formation process remain largely unresolved, it is generally believed that large spiral galaxies like the Milky Way have been assembled from numerous low- to intermediate-mass clumps of stars and gas, each of which is found in its own dark matter sub-halos, through mergers and accretions onto more massive halos. The emerging picture, supported by a number of recent large cosmological simulations \citep[e.g., ][]{zolotov:09, font:11, tissera:13, tissera:14}, is a generic expectation that the metallicity distribution functions (MDFs), kinematics, and spatial distributions of the stellar populations formed primarily from early mergers and in situ star formation differ from those that formed primarily from accretion. 

In the past decade our view of the nature of the halo of the Galaxy has also evolved substantially, with considerable observational support for the suggestion that it comprises at least two distinct diffuse stellar components, the inner-halo and outer-halo populations \citep[][and references therein]{carollo:07,carollo:10, dejong:10, nissen:10,beers:12, schuster:12,an:13}, with substantial contributions from the debris of relatively recent accretion events (such as the Sagittarius Stream) in the outer regions of the halo \citep{janesh:15}, in contrast to the previously assumed monolithic halo. In any event, it is clear that full understanding of the complex nature of the Milky Way's halo system is required in order to constrain models for the structural and chemical evolution of large spirals like our own.

The first inferences for a dual halo relied on a sample of stars exploring only a nearby region ($d_{\rm sun} \le 4$ kpc), so that accurate proper motions could be used \citep{carollo:07,carollo:10}.  This led to criticism (e.g., \citealt{schonrich:11}; refuted by \citealt{beers:12}) that the derived space motions were affected by distance-scale errors, creating the illusion of a dual halo.  Interpretations based on more distant in situ samples of stars, exploring regions many tens of kpc away \citep{chen:14, allende:14, fernandez-alvar:15}, have now firmly established the presence of trends in the mean metallicities, as well as in individual abundance ratios ([Ca/Fe], [Mg/Fe]), indicating that the diffuse halo outside the local region behaves in essentially the manner required by the initial claims of \citeauthor{carollo:07} and others. The apparent differences in the frequencies of the sub-classes of carbon-enhanced metal-poor \citep[CEMP; see][]{beers:05} stars that can be associated with the inner- and outer-halo populations are also difficult to reconcile with a single-halo model \citep{carollo:12, carollo:14}. 

Another limiting factor in consideration of the nature of the halo is that it has proven extremely challenging to assemble what might be referred to as a ``fair'' sample of stellar probes. ``Historical samples,'' going back to selections based on high proper motions \citep[e.g.,][]{ryan:91} necessarily confounded kinematics with chemistry, under the assumption that this would not severely impact interpretations of the MDF from a single-halo population. Others approached the problem by selecting samples of halo stars based on the apparent weakness of metallic lines (in particular \ion{Ca}{2} K) from objective-prism surveys, which necessarily distorts the MDF of the more metal-rich stars in favor of producing samples of the most metal-poor stars \citep{beers:05, frebel:15}.

An alternative approach to the above was first explored by \citet{ivezic:08a}, who produced a photometric-metallicity map based on estimates from SDSS $ugr$ photometry, although it was acknowledged that the techniques employed failed for metallicities [Fe/H] $< -2.0$, so tests for the presence of a very low-metallicity outer-halo population could not be made.

\citet[][hereafter A13]{an:13} explored a refined approach to the photometric metallicity-estimation technique, based on comparisons of $ugriz$ photometry with well-calibrated cluster fiducial sequences. For a sample of several thousand stars from SDSS Stripe 82, selected to span a limited range of main-sequence mass and apparent magnitude (resulting in distance limits 5-8 kpc from the Sun), A13 were able to obtain the first fair sample of local halo stars, one that is unbiased in {\it both} kinematics and chemistry (covering the metallicity range $-2.5 \le {\rm [Fe/H]} \le -1.0$, and perhaps a little lower). The relatively limited number of stars in their final sample ($N\sim2,500$) and the influence of photometric errors on metallicity estimates at the low end of the MDF prevented them from confirming the presence of a dual halo from the derived MDF alone, as compared to a ``simple model'' distribution \citep[e.g.,][]{hartwick:76,norris:91} -- the data were statistically consistent with being drawn from either model of the parent population.  However, A13 {\it were} able to demonstrate, on the basis of an approximate rotational measure (derived from proper motions of stars at high Galactic latitudes) that stars with retrograde velocities clearly favored the lowest metallicity stars in their sample, and concluded that a single halo population was inconsistent with this result.

Recently, \citet[][hereafter J14]{jiang:14} revisited the SDSS Stripe 82 $ugriz$ data, producing a ``depth optimized'' co-add sample with fainter limiting magnitudes and reduced errors in the final photometry.  \citet{yuan:15b} used these data to assemble a catalog of photometric metallicity estimates for a half million stars, spanning a large color and metallicity range. It appears, however, that their method of metallicity estimation suffers from similar limitations at low metallicity as did the \citet{ivezic:08a} approach, rendering it unsuitable for exploration of the halo MDF for [Fe/H] $\la -2.0$, where the potential contribution from outer-halo stars becomes dominant.  Hence, in the present paper we have used the approach of A13 and apply it to the J14  co-adds, in order to explore anew the nature of the derived halo MDF with a substantially larger sample of stars with improved photometric metallicity estimates.

\begin{figure}
\centering
\includegraphics[scale=0.45]{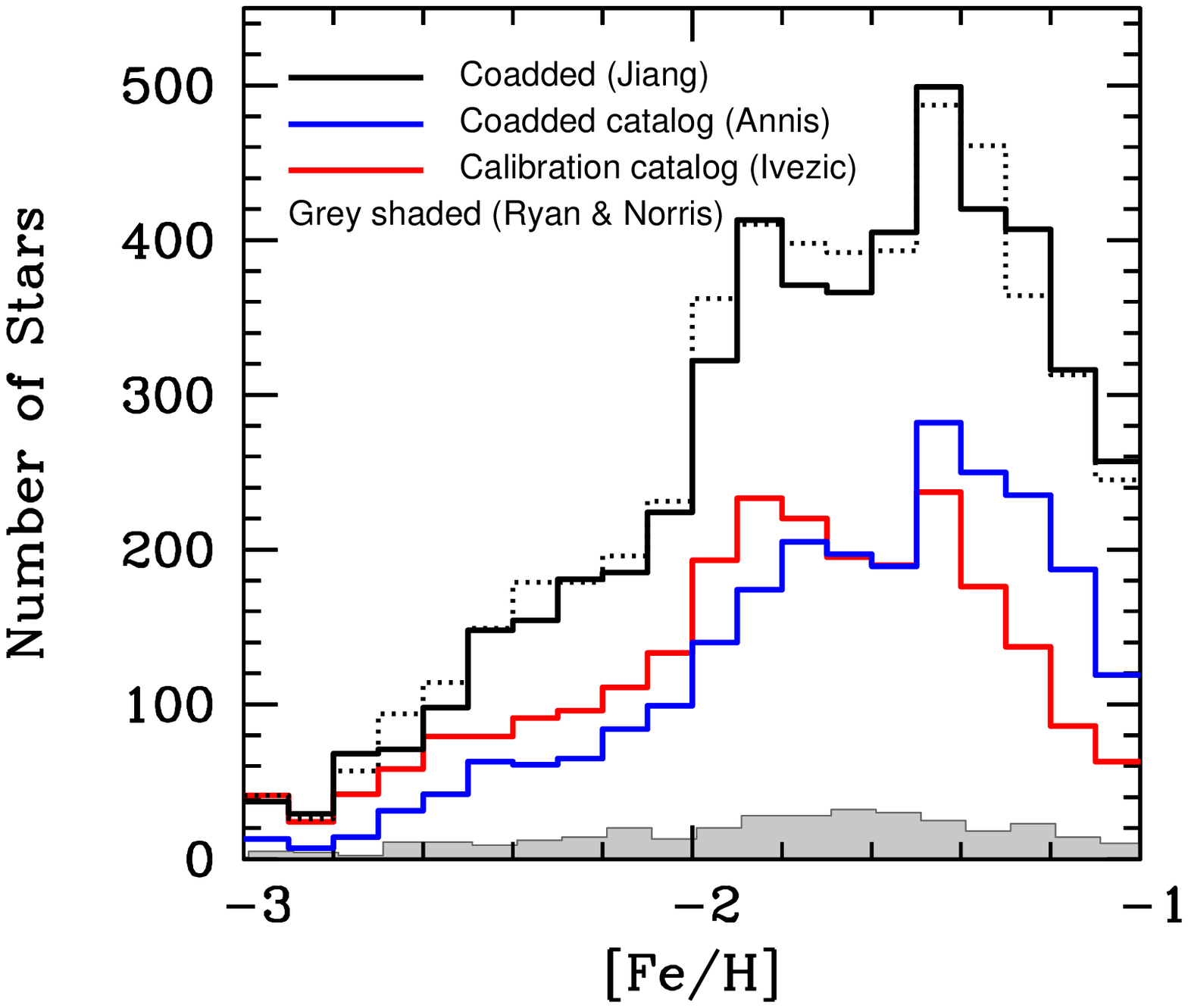}
\includegraphics[scale=0.45]{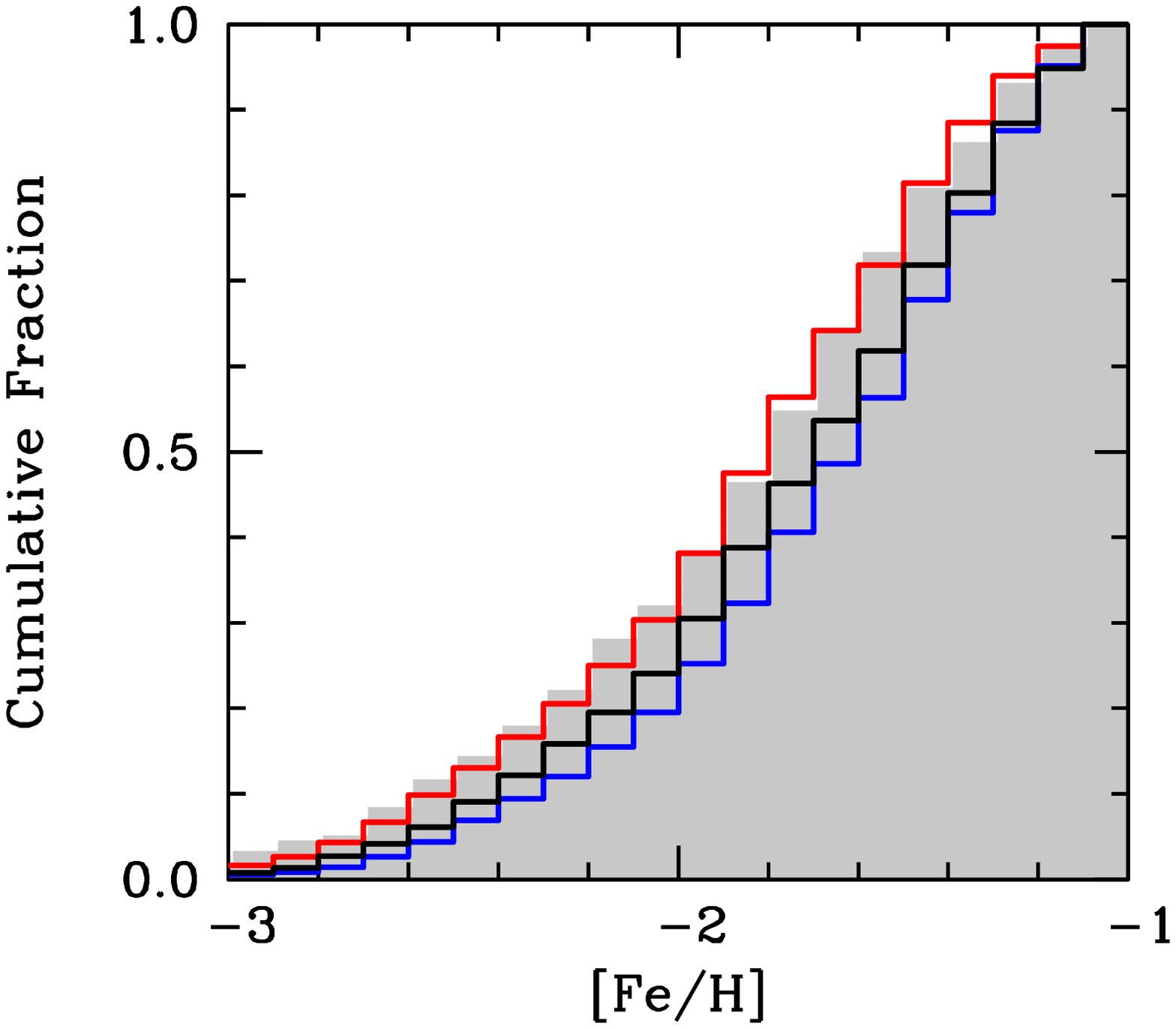}
\caption{{\it Top panel:} comparison of the new MDF from the J14 (black solid line) co-add catalog with those derived in A13 using the calibration \citep[][red solid line]{ivezic:07} and co-add \citep[][blue solid line]{annis:14} catalogs. The gray region is the MDF from the spectroscopic sample of \citet{ryan:91}. The black dotted line shows the MDF derived from the J14 catalog, but after applying photometric zero-point corrections found in \citet{yuan:15a}. {\it Bottom panel:} cumulative MDFs for the above samples.  Note that, although there are more stars included in the newer catalog, the overall shapes of MDF are similar to one another.  \label{fig:mdf}} \end{figure}

\section{The Updated SDSS Co-add Photometry}

We employ $ugriz$ photometry for Stripe~82 from the co-added imaging frames in J14, a catalog that is almost two magnitudes deeper than the single-epoch photometry in SDSS. Compared to \citet{annis:14}, the photometry is also deeper by $0.3$--$0.5$~mag, due to the use of more repeated image scans in the co-add procedure. Following the prescription in J14, we used aperture photometry with $8$ pixel ($3.2\arcsec$) diameter aperture, and applied aperture corrections based on the difference from the photometry with $20$ pixel ($8.0\arcsec$) aperture for relatively bright stars ($15 < r < 18$) in each field. We selected good photometry based on the {\tt SExtractor} extraction flag set to zero in each band. We then assembled matched photometry in five filter passbands using a $1\arcsec$ search radius.

\section{Sample Selection}

Figure~$12$ of A13 illustrates our selection of halo main-sequence stars.  Higher-mass main-sequence stars can be used to probe a more distant volume than lower-mass stars, because they are intrinsically brighter than less-massive ones. The mass-luminosity relations also depend strongly on metallicity, so metal-poor stars are seen at larger distances at a given stellar mass. These relatively well-known relations between stellar mass, luminosity, and metallicity set a strict constraint on the sample selection from a large photometric database, if one desires to obtain an unbiased sample against metallicity. Unfortunately, there is no adequate distance-mass bin that is essentially bias-free over a large range in metallicity. Instead, we focused our analysis on metal-poor stars ([Fe/H]$<-1.2$) to construct an unbiased MDF of the halo system, as in A13.

Photometric metallicities are sensitive to the size of photometric errors in the $u$ passband, because the technique relies on strong metallic-line blanketing in this short wavelength band. In A13, we used photometry of stars with $\sigma_u < 0.03$~mag, or $u < 20.6$~mag and $u < 21.0$~mag in the calibration \citep{ivezic:07} and co-added \citep{annis:14} catalogs, corresponding to maximum heliocentric distances of $\sim7.8$~kpc and $\sim9.4$~kpc, respectively. With the new co-add catalog of J14, we can extend our photometric metallicity estimates of stars out to $\sim10.5$~kpc from the Sun, by virtue of its fainter $u$-band limit ($\sigma_u=0.03$~mag at $u=21.3$~mag).

Other sample-selection criteria are the same as in A13: stars must be detected in all five $ugriz$ pass bands, with a reduced $\chi^2$ value of a model fit less than 3 for each star, and located at high Galactic latitudes ($|b| > 35\arcdeg$), with a minimum heliocentric distance of $5$~kpc in order to minimize contamination from disk-system stars. A minimum mass of $0.65\ M_\odot$ was set by the minimum heliocentric distance and the above mass-luminosity-metallicity limit.  The maximum stellar mass was set to $0.75\ M_\odot$, since stars with $M_* \sim 0.75\ M_\odot$ are near the main-sequence turn-off at the lower end of our metallicity estimate ([Fe/H] $ = -3$).

\section{Updated Halo Metallicity Distribution Functions and Metallicity-Kinematics Correlations}

Our updated MDF of the halo is presented as a black histogram in Figure~\ref{fig:mdf} (see also Table~\ref{tab:tab1}), which comprises a stellar sample that is almost a factor of two larger ($N=4971$ at [Fe/H]$<-1$) than those of the \citet[][red histogram; $N=2484$]{ivezic:07} or the \citet[][blue histogram; $N=2457$]{annis:14} catalogs used in our previous effort. All three SDSS catalogs are enormously larger than the kinematically-selected stars in the solar neighborhood from \citet{ryan:91}, for which the MDF is shown in the gray shaded region. The black dotted histogram in the top panel shows the same MDF as the black solid histogram, but after applying photometric zero-point corrections as proposed by \citet{yuan:15a} in both the scan and camcol directions; the differences are negligible. The bottom panel of Figure~\ref{fig:mdf} shows cumulative distributions of these MDFs; the shape of our new MDF is similar to those from the previous samples.

\begin{deluxetable}{rrr}
\tablewidth{0pt}
\tablecaption{Photometric Metallicity Distribution Function from Stripe~82\label{tab:tab1}}
\tablehead{
  \colhead{{\rm [Fe/H]}} &
  \colhead{$N_{\rm stars}$} &
  \colhead{$N_{\rm stars}$ (corr)\tablenotemark{a}}
}
\startdata
$-$1.05 & 257 & 245 \nl
$-$1.15 & 316 & 313 \nl
$-$1.25 & 407 & 364 \nl
$-$1.35 & 420 & 461 \nl
$-$1.45 & 499 & 487 \nl
$-$1.55 & 405 & 393 \nl
$-$1.65 & 366 & 392 \nl
$-$1.75 & 371 & 398 \nl
$-$1.85 & 413 & 410 \nl
$-$1.95 & 322 & 362 \nl
$-$2.05 & 224 & 231 \nl
$-$2.15 & 185 & 196 \nl
$-$2.25 & 181 & 179 \nl
$-$2.35 & 154 & 179 \nl
$-$2.45 & 148 & 149 \nl
$-$2.55 &  98 & 114 \nl
$-$2.65 &  71 &  94 \nl
$-$2.75 &  68 &  57 \nl
$-$2.85 &  29 &  26 \nl
$-$2.95 &  37 &  41 \nl
\enddata
\tablecomments{Our sample selection is biased against stars with ${\rm [Fe/H]} >-1.2$ (see A13).}
\tablenotetext{a}{After applying photometric zero-point corrections in \citet{yuan:15a}.}
\end{deluxetable}

\begin{figure}
\epsscale{1.1}
\plotone{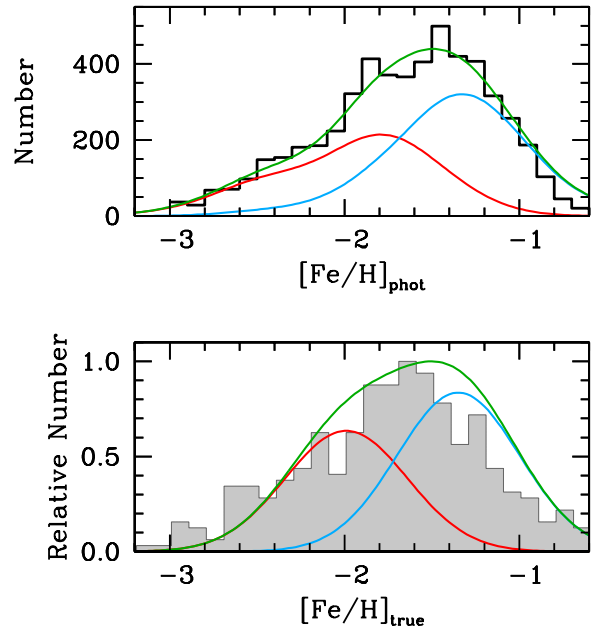}
\caption{{\it Top panel}: the observed photometric MDF of the new co-added sample (black histogram), and its deconvolution using two components (red and blue lines). Each curve represents a MDF with a Gaussian metallicity distribution (shown in the bottom panel), convolved with error+binary models as described in A13. The green curve is a sum of these two components.  {\it Bottom panel:} the modeled input components (red and blues lines, with dispersions of $0.4$~dex, respectively) and the resulting mixture (green line).  The MDF in the solar neighborhood from \citet{ryan:91} is shown as a gray histogram for comparison.  \label{fig:decomp}} \end{figure}

Figure~\ref{fig:decomp} is a decomposition of the newly-determined MDF, under the assumption that the diffuse halo system is composed of two spatially overlapping populations of stars. Specifically, we adopted the working hypothesis that these MDFs can be described as Gaussian distributions in [Fe/H], with a dispersion of $\sigma = 0.4$~dex for each component. As in A13, we did not employ the simple mass-loss modified model in a leaky box \citep[e.g.,][]{hartwick:76}, since such models are based on unphysical (and unproven) assumptions on the instantaneous recycling and mixing of metals, which are incompatible with a presumed hierarchical assembly of the Galactic halo from individual lower-mass subhalos. The red and blue lines in the top panel show the division of the observed MDF, which take into account the smearing due to photometric errors and unresolved binaries in the sample. The green line is a sum of these two MDFs. The peak of the (deconvolved) metal-rich component is [Fe/H]$=-1.38\pm0.26$, while that of the metal-poor component is [Fe/H]$=-1.94\pm0.29$, with little dependence on the choice of the dispersion in [Fe/H]. Our estimated metallicity peaks are systematically $\sim0.2$~dex higher than the (spectroscopic) values ([Fe/H]$=-1.6$ and $-2.2$, respectively) found by \citet{carollo:07}.

\begin{figure*}
\epsscale{0.95}
\plotone{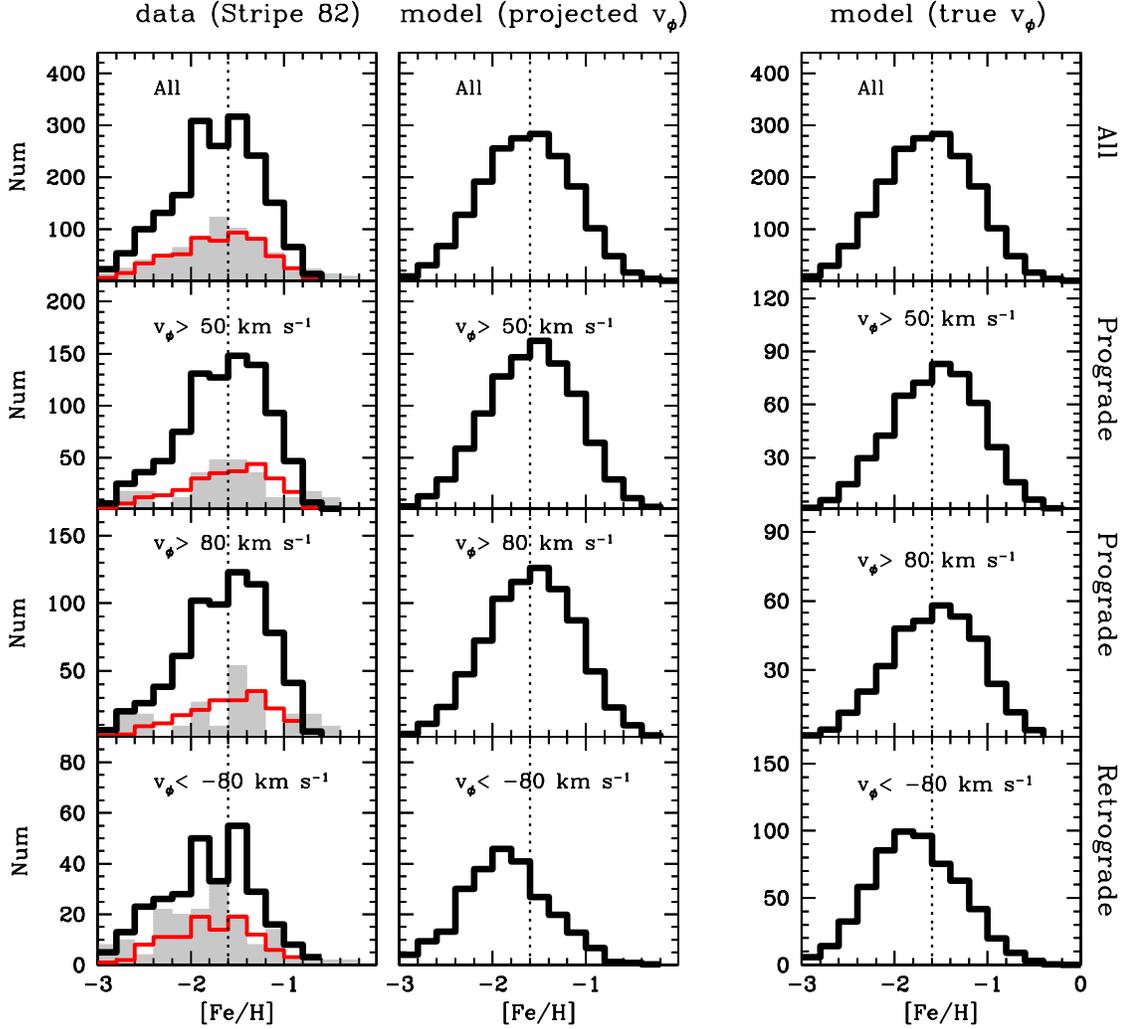}
\caption{{\it Left column:} changes in the observed MDF of local halo-system stars for different cuts in $v_\phi$. The black histogram shows MDFs of stars at $|b| > 45\arcdeg$, and the red histogram shows those with $|b| > 60\arcdeg$.  The gray histograms display local MDFs from \citet{ryan:91}, after multiplying their MDF by an arbitrary factor to approximately match the histograms of the high-latitude sample. There is a clear preference for low-metallicity stars to be found in greater number, relative to metal-rich stars, in the retrograde cuts compared to the prograde cuts. {\it Middle column:} halo MDFs as derived from the best-matching Milky Way models, assuming a fractional contribution of $55\%$ from inner-halo stars in our sample volume. The same sample cuts in $v_\phi$ (and $|b| > 45\arcdeg$) are shown as in the left column, but $v_\phi$ is derived from proper motions in the mock catalog. {\it Right column:} same as in the middle panels, but based on $v_\phi$ derived from full kinematic information in the mock catalog.\label{fig:feh}} \end{figure*}

About $10\%$ of the stars in our sample are found at [Fe/H]$ < -2.3$, as is also the case for the \citeauthor{ryan:91} sample. The estimated fraction of the metal-poor component amounts to $\sim40\%$--$45\%$, depending on the assumed size of a dispersion in [Fe/H] for each component ($\sigma\sim0.2$--$0.4$~dex). When the peak metallicity of the inner-halo component changes by $\pm0.1$~dex, the peak metallicity of the outer-halo component also changes by approximately $0.1$~dex (positively correlated). The fraction of inner/outer halo increases/decreases by $\sim15\%$ when the peak metallicities decreases by $0.1$~dex. Interestingly, the fractional contribution from the metal-poor component is larger than that found in A13 ($\sim30\%$), perhaps because we included more distant stars in our sample from the J14 catalog.

\begin{figure*}
\epsscale{0.95}
\plotone{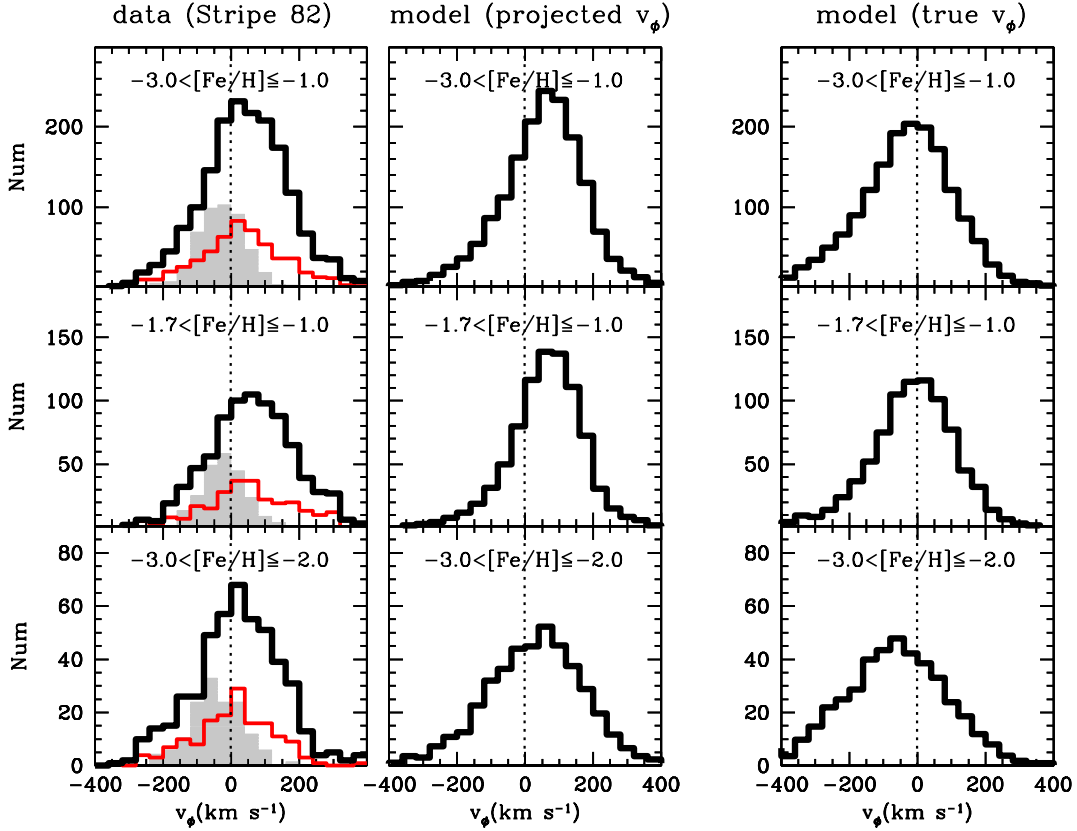}
\caption{Same as in Figure~\ref{fig:feh}, but showing the $v_\phi$ distribution for different sample cuts in metallicity.  \label{fig:vphi}} \end{figure*}

In order to inspect the properties of the two chemically- and kinematically-distinct populations, we combined our photometric metallicity estimates with approximate Galactocentric rotational velocities ($v_\phi$). While a $v_\phi$ determination requires full kinematic information from both radial-velocity and proper-motion measurements, proper motions alone (combined with distances) can be used to derive approximate rotational velocities for stars at high Galactic latitudes. The left-hand column in Figure~\ref{fig:feh} shows MDFs from our new co-add sample. The black histograms are MDFs of stars at $|b| > 45\arcdeg$, with good proper-motion measurements from \citet{munn:04}.  The red histograms are stars with $|b| > 60\arcdeg$, which should yield more accurate estimates of $v_\phi$. For comparison, the grey shaded histogram represents the MDF from \citet{ryan:91} based on full space motions.

In the lower rows of panels shown in Figure~\ref{fig:feh}, we divided our sample stars based on their rotational velocities, either prograde ($v_\phi > 50$~km s$^{-1}$ or $v_\phi > 80$~km s$^{-1}$)\footnote{The $v_\phi > 50$~km s$^{-1}$ cut in the second row of panels was added due to the small number of stars in the \cite{ryan:91} sample with $v_\phi > 80$~km s$^{-1}$.} or retrograde ($v_\phi < -80$~km s$^{-1}$). As can be seen from comparison of the lower two panels in the left-hand column, there is a clear shift in the MDFs for stars in the prograde and retrograde sub-samples. The median metallicity of stars ($|b| > 45\arcdeg$) with prograde rotations (third panel) is [Fe/H]$=-1.6$, while that of stars with retrograde rotations (bottom panel) is [Fe/H]$=-1.8$. A Kolmogorov-Smirnov (K-S) test rejects a null hypothesis that the two distributions were drawn from the same underlying population at high levels of significance ($p=0.00004$ for $|b| \geq 45\arcdeg$; $p=0.007$ for $|b| \geq 60\arcdeg$). These results are even more statistically significant than found by A13 for the \citet{annis:14} co-add catalog, due to the factor of two increase in the number of stars included in the halo sample, suggesting that the Milky Way halo system comprises at least two stellar populations with distinct metallicity-kinematics properties \citep[e.g.,][]{carollo:07,carollo:10,beers:12}. Note that similar differences are seen for the \citet{ryan:91} solar-neighborhood sample (which has been rescaled to match the numbers found for our present high-latitude sample).

Similarly, the left-hand column in Figure~\ref{fig:vphi} shows the distribution of $v_\phi$ for stars in Stripe~82. The top panel in this column contains our sample stars with photometric metallicities $-3 < {\rm [Fe/H]} \leq -1$. Stars in the middle panel are stars with higher metallicities ($-1.7 < {\rm [Fe/H]} \leq -1$) than those shown in the bottom panel ($-3 < {\rm [Fe/H]} \leq -2$).  Comparison between the lower two panels clearly shows that high-metallicity halo stars are, on average, rotating faster in the same direction as the disk ($\langle v_\phi \rangle = +61\pm5$~km s$^{-1}$, while those with lower metallicities are rotating more slowly ($\langle v_\phi \rangle = +16\pm6$~km s$^{-1}$). A K-S test rejects a null hypothesis that the two distributions were drawn from the same underlying population at high levels of significance ($p=10^{-6}$ for $|b| \geq 45\arcdeg$; $p=0.006$ for $|b| \geq 60\arcdeg$).

To provide a check on our inferences, we examine the behavior of simulated samples of main-sequence stars selected from mock catalogs created by running {\tt galfast} \citep{juric:08,juric:10}. We considered chemically- and kinematically-distinct stellar components in the halo with fixed ratios of contributions, ranging from $10\%$ to $90\%$ of inner-halo stars (90\% to 10\% outer-halo stars), with a $10\%$ increment. Instead of deriving photometric metallicities of stars from synthetic color-magnitude diagrams in the mock catalog, we directly assigned stellar metallicities by drawing from parent Gaussian metallicity distributions with peaks at [Fe/H]$=-1.38$ and $-1.94$ for the inner and outer halos, respectively, each of which has a metallicity dispersion of $\sigma = 0.4$~dex. We adopted the parameters of the velocity ellipsoids (and derived power-law slopes) in the halo system from \citet{carollo:10}, where the mean rotational velocities of the two components are $\langle v_\phi \rangle = +7$~km s$^{-1}$ and $-80$~km s$^{-1}$, respectively (when $v_{\phi,\rm LSR}=+220$~km s$^{-1}$), and power-law slopes of $-3.17$ and $-1.79$ for the inner- and outer-halo density profiles, respectively. We generated mock catalogs of stars along the Stripe~82 footprint with the same limit in Galactic latitude ($|b|>45\arcdeg$). To match our mass-luminosity-metallicity criterion, we restricted simulated stars to $4 \leq M_r \leq 6$ (see Figure~11 in A13), along with $5 \leq d_{\rm sun} \leq 10$~kpc.

The middle and right-hand columns in Figures~\ref{fig:feh} and \ref{fig:vphi} show the predicted MDFs and $v_\phi$ distributions from {\tt galfast}, when the modeled contribution from the inner halo is set to $55\%$ in our sample volume.  Each panel shows a distribution of stars with the same $v_\phi$ (or [Fe/H]) cut as in the left-hand column. In the middle column, we used a ``projected $v_\phi$,'' computed from proper motions and distances as in our observed dataset. Since our Stripe~82 sample extends over $60\arcdeg \la l \la 180\arcdeg$ and $-65\arcdeg \la b \la -45\arcdeg$, these projected $v_\phi$ values are systematically higher than the true rotation velocities based on full kinematic information from the mock catalog (right-hand column).

As shown in the lower panels in the middle column of Figure~\ref{fig:feh}, a shift in the peak metallicity (and tails) of the simulated dataset is clearly present, similar in nature to the observed dataset, when divided into stars with prograde and retrograde orbits. Each subsample includes both inner- and outer-halo stars, because of their large velocity dispersions, but the shift in the overall MDFs can be understood as due to the different contributions from each of the components in the $v_\phi$ cuts. We looked for the best-matching fraction of inner/outer halos in the models by changing their relative contributions, and found that $\sim45\%$--$65\%$ inner-halo fractions provide acceptable matches to the observed MDFs and $v_\phi$ distributions in Figures~\ref{fig:feh} and \ref{fig:vphi}.  Lower (or higher) inner-halo fractions result in a significant enhancement (or depression) of the outer-halo component, yielding incompatible matches to these distributions. In all cases, a single-population diffuse-halo model is rejected. 

\section{Conclusions}

In this letter we have employed the deep photometric co-add catalog of Stripe~82 (J14) to derive new estimates of photometric metallicities for halo main-sequence stars, and obtained a fair MDF based on more distant in situ samples of stars than available in our previous work. Our results show that stars with retrograde orbits have, on average, lower metallicities than stars with prograde rotations, consistent with previous results from A13, but with even higher statistical significance. The estimated fraction of the metal-poor or outer-halo component in the local volume is $\sim45\%\pm10\%$.

The presence of two distinct diffuse stellar halos implies that the formation of the halo system must have involved at least two different star-formation episodes. According to recent theoretical models \citep[e.g.,][]{johnston:08,zolotov:09,mccarthy:12,tissera:13, tissera:14}, inner-halo population stars formed from more massive sub-halos, or dwarf galaxies with sufficient gas content that could reach the inner region of the proto-Galaxy through dynamical friction, or formed in situ in the inner region from the rapid collapse of primordial gas. Sustained star formation in this region would lead to higher metallicity. In contrast, stars of the outer-halo population formed in lower mass dwarf galaxies, and were brought into the main halo of the Milky Way through disruption and accretion, resulting in a diffuse outer halo with distinct and significantly hotter kinematics, which is present throughout the halo system of the Galaxy, including the inner-halo region. A limited star-formation history, perhaps involving no more than a single burst, would lead to lower metallicity. The situation is likely to be complex, and dependent on both the star-formation history within individual sub-halos, and on their merger and accretion histories (see discussions by \citealt{johnston:08} and \citealt{tissera:13}). In any event, one would expect to find samples of old stars arising from both populations in the local volume.  This interpretation has received recent support from the identification by \citet{santucci:15} of the so-called ``ancient chronographic sphere'', a region comprising the oldest stars in the Milky Way located within 15 kpc from the Galactic center, and extending into the region we have analyzed in the present paper.

Our work is limited by the photometric precision and kinematic measurement errors in the currently available survey data, as well as the relatively small volume of space explored by Stripe~82. However, it clearly demonstrates the feasibility of dissecting stellar populations in more distant regions of the halo with future large photometric and astrometric surveys such as the Large Synoptic Sky Survey \citep[LSST;][]{ivezic:08b} and Gaia \citep{perryman:01}.

\acknowledgements

We thank Mario Juri{\'c} for his help with {\tt galfast}. D.A.\ and Y.S.L.\ acknowledge support provided by the National Research Foundation of Korea to the Center for Galaxy Evolution Research (No.\ 2010-0027910). D.A.\ was partially supported by Basic Science Research Program through the National Research Foundation of Korea (NRF) funded by the Ministry of Education (2010-0025122). T.C.B., V.M.P., \& D.C.\ acknowledge partial support for this work from grant PHY 14-30152; Physics Frontier Center/JINA Center for the Evolution of the Elements (JINA-CEE), awarded by the US National Science Foundation.  R.M.S.\ and S.R.\ would like to thank partial support from FAPESP, CNPq, and CAPES.  Y.S.L.\ acknowledges partial support from the Basic Science Research Program through the National Research Foundation of Korea (NRF) funded by the Ministry of Science, ICT \& Future Planning (NRF-2015R1C1A1A02036658).

{}

\end{document}